\documentclass[12pt,a4paper]{article}

\usepackage{latexsym,amsmath,amssymb}
\allowdisplaybreaks
\renewcommand{\thefootnote}{\fnsymbol{footnote}}
\newcommand{\acknowledgments}{%\vspace{1cm}
$\\${\bf Acknowledgments}\newline}

\begin{document}
\begin{flushright}
%{\rm arXiv:yymm.nnnn}
%{\rm CQUeST-2011-xxxx}
\end{flushright}
\vspace{12mm}
\begin{center}
{{{\Large {\bf Area law of the entropy in the critical gravity}}}}\\[10mm]
{Edwin J. Son$^{a,b}$\footnote{email:eddy@sogang.ac.kr},
Myungseok Eune$^{c}$\footnote{email:younms@sogang.ac.kr}, and
Wontae Kim$^{a,b}$\footnote{email:wtkim@sogang.ac.kr}}\\[8mm]

{{${}^{a}$ Department of Physics, Sogang University, Seoul, 121-742, Korea\\[0pt]       
${}^{b}$ Center for Quantum Spacetime, Sogang University, Seoul, 121-742, Korea\\[0pt]
${}^{c}$ Research Institute for Basic Science, Sogang University, Seoul, 121-742, Korea}\\[0pt]
}
\end{center}
\vspace{2mm}
\begin{abstract}
The entropy of the  Schwarzschild-anti de Sitter black hole in the recently proposed four-dimensional critical gravity
is trivial in the Euclidean action formulation, while it is expressed by the area law in terms of the
brick wall method given by 't~Hooft. To resolve this issue, 
we relate the Euclidean action formulation to the brick wall method semiclassically, and 
show that the entropy of the black hole 
can be expressed by the area law at the critical point.
\end{abstract}
\vspace{5mm}

%{\footnotesize ~~~~PACS numbers: 04.70.Dy, 04.60.Kz}


\vspace{1.5cm}

\hspace{11.5cm}{Typeset Using \LaTeX}
\newpage
\renewcommand{\thefootnote}{\arabic{footnote}}
\setcounter{footnote}{0}
%%%%%%%%%%%%%
%% LaTeX Style End %%
%%%%%%%%%%%%%

\section{Introduction}
\label{sec:intro}

Since it has been claimed that the general relativity is nonrenormalizable,
there have been extensive studies for quantum theory of gravity such as
string theory, conventional perturbative gravity, and so on. 
In particular, one of the perturbatively renormalizable gravity theories can be built by adding
quadratic curvature terms to the Einstein
gravity~\cite{Stelle:1976gc,Stelle:1977ry}. 
However, theories including
higher-order time-derivative terms should endure massive ghost modes. In
recent studies on the three-dimensional topologically massive
gravity~\cite{Deser:1981wh,Deser:1982vy} including a cosmological
constant, it has been shown that there exists some critical point such that the massive mode
becomes massless and carries no energy, so that the problem can be
solved~\cite{Li:2008dq}. 

Similarly, in the four-dimensional
quadratic gravity theory with a cosmological constant, one can find a critical
point, where the massive ghost mode disappears. This model is called the
\emph{critical gravity}~\cite{Lu:2011zk} defined by
\begin{align}
  I_\text{CG} [g] &= \frac{1}{16\pi G} \int d^4x \sqrt{-g} \left[ R - 2\Lambda + \alpha \left( R_{\mu\nu} R^{\mu\nu} - \frac13 R^2 \right) \right] \label{action:CG}
\end{align}
with a cosmological constant $\Lambda$.
At the critical point, $\alpha = 3/2\Lambda$, in spite of the renormalizability, 
it seems to be trivial in the sense that 
the mass and the entropy of a Schwarzschild-anti de Sitter
(SAdS) black hole which is a solution to this theory become zeroes~\cite{Lu:2011zk}.
Moreover, this result can be also confirmed by the Euclidean action formulation of
the black hole thermodynamics~\cite{Gibbons:1976ue,Gibbons:1978ac,Hawking:1982dh}.

On the other hand, it has been suggested by Bekenstein that the
intrinsic entropy of a black hole is proportional to the surface area
at the event horizon~\cite{Bekenstein:1972tm, Bekenstein:1973ur,
  Bekenstein:1974ax}, and then quantum field theoretic
calculation has been given for the Schwarzschild black
hole by Hawking~\cite{Hawking:1974sw}.
Actually, one of the best way to reproduce the area law of black holes is to use
the brick wall method suggested by 't~Hooft~\cite{thooft}. 
By considering fluctuation of matter field around black holes semiclassically, one can always get the 
desired results, however, this result is not compatible with the result of the Euclidean action formulation
for the critical gravity.  

In this paper, we would like to resolve  
the above-mentioned issue and study how to derive the entropy  satisfying the area law
in terms of the Euclidean action formulation. First task is to get  the nontrivial free energy by taking
into account higher-order corrections in the Euclidean path integral and then 
the corresponding entropy can be nontrivial.
For convenience, the fluctuation of the metric field will be ignored, i.e. our
calculations will be performed in semiclassical approximations.
We recapitulate the Euclidean action formulation
by carefully considering the appropriate boundary term
in section~\ref{sec:Euclideanaction}. In contrast to conventional cases, 
the entropy is trivially zero assuming the critical condition.
It means that the partition function is trivial so that the area law of the entropy does not appear. 
So, we consider the one loop correction of the scalar degrees of freedom around the black hole
in section~\ref{sec:pathintegral} and relate the Euclidean action formulation to the brick wall method 
semiclassically. Eventually, the free energy turns out to be nontrivial even at the
critical condition where it is actually compatible with the free energy from the brick wall method with
some conditions, which yields the area law of the black hole entropy in
section~\ref{sec:thermo}. 
Finally, in section~\ref{sec:dis}, summary and some discussions are given.

\section{Thermodynamics with Euclidean action formulation}
\label{sec:Euclideanaction}
We start with a minimally coupled scalar field $\phi$ coupled to the critical gravity as $I_\text{tot} =
I_\text{CG}+I_\phi$, where the scalar field action is 
\begin{align}
  I_\phi [g, \phi] &= - \int d^4 x \sqrt{-g} \left[ \frac12 (\nabla \phi)^2 +
    \frac12 m^2 \phi^2 \right]. \label{action:scalar}
\end{align}
For $\phi=0$, the SAdS black hole is just a classical solution to this model.
The line element of the SAdS black hole is given by $ds^2  = g_{\mu\nu}
dx^\mu dx^\nu = - f dt^2 + f^{-1} dr^2 + r^2 d\Omega_2^2$ with
\begin{align}
  f(r) = 1- \frac{2GM}{r} - \frac{\Lambda}{3}  r^2= \left(1 -
    \frac{r_h}{r} \right) \left[1 - \frac{\Lambda}{3} (r^2 + r_h r +
    r_h^2) \right], \label{f:SAdS}
\end{align}
where $M = (r_h/2G) \left( 1 - \Lambda r_h^2/3 \right) > 0$ is the mass
parameter of the black hole, $\Lambda<0$ is the cosmological constant,
and $r_h$ is the radius of the horizon.
The free energy $F^{(0)}$ for this vanishing scalar solution
can be obtained from Euclidean action formulation~\cite{Gibbons:1976ue,Gibbons:1978ac,Hawking:1982dh},
\begin{equation}
  Z^{(0)} [g] = \exp \left( i I_\text{CG} [g] \right)
  = \exp \left( -\beta F^{(0)} \right).
\end{equation}
The crucial ingredient for this calculation is to find the consistent boundary term.
Following ref.~\cite{Hohm:2010jc}, an auxiliary field $f_{\mu\nu}$ is
introduced to localize the higher curvature terms so that the Euclidean version of the action~\eqref{action:CG}
and the corresponding boundary term can be written in the form of
\begin{align}
  I_\text{CG} &= -\frac{1}{16\pi G} \int_{\mathcal{M}} d^4x \sqrt{g} \left[ R - 2\Lambda + f^{\mu\nu} \left( R_{\mu\nu} - \frac12 g_{\mu\nu} R \right) 
    - \frac{1}{4\alpha} f^{\mu\nu} \left( f_{\mu\nu} - g_{\mu\nu} f \right) \right], \\
  I_\text{B} &= -\frac{1}{16\pi G} \int_{\partial \mathcal{M}} d^3x \sqrt{\gamma} \left[ 2 K + \hat{f}^{ij} \left( K_{ij} - \gamma_{ij} K \right) \right],
\end{align}
where $\gamma_{ij}$ and $K_{ij}$ are the induced metric and the
extrinsic curvature of the boundary, respectively. And $\hat{f}^{ij}$
in the boundary term is defined as $\hat{f}^{ij} = f^{ij} + f^{ri} N^{j} + 
f^{rj} N^{i} + f^{rr} N^{i} N^{j}$ with $N^{i} = - g^{ri}/g^{rr}$ for the
hypersurface described by $r=r_0$.
In the Euclidean
geometry, the Euclidean time is defined by $\tau = i t$ and 
should be identified by $\tau=\tau+\beta_H$ to avoid a conical singularity at the event horizon,
where $\beta_H$ is the inverse of the Hawking temperature.

Next, taking the boundary to the infinity, $r_0\to\infty$, the free energy is obtained as
\begin{align}
  F^{(0)} &= \beta_H^{-1} (I - I_\text{vacuum}) %\notag \\ &
  = \left[ 1-2\alpha\Lambda/3 %1 + 2\Lambda \left( \alpha + 4\beta \right) 
  \right] \frac{r_h}{4G} \left( 1 + \frac{\Lambda}{3} r_h^2 \right), \label{F:tree}
\end{align}
where $I = I_\text{CG} + I_\text{B}$ and $I_\text{vacuum} = I|_{M=0}$,
and the Hawking temperature for the given metric function~\eqref{f:SAdS} is
calculated as
\begin{equation}
  \label{T:H}
  T_H = \beta_H^{-1} = \frac{1-\Lambda r_h^2}{4\pi r_h}.
\end{equation}
Then, the thermodynamic first law reads the entropy and the energy of the black hole,
\begin{align}
  S^{(0)} &= \beta_H^2 \frac{\partial F^{(0)}}{\partial \beta_H} %\notag \\ &
  = \left[ 1-2\alpha\Lambda/3 %1 + 2\Lambda \left( \alpha + 4\beta \right) 
  \right] \frac{\pi r_h^2}{G}, \label{S:tree} \\
%\end{align}
%and the energy of the black hole is obtained as
%\begin{align}
  E^{(0)} &= F^{(0)} + \beta_H^{-1} S^{(0)} %\notag \\ &
  = \left[ 1-2\alpha\Lambda/3 %1 + 2\Lambda \left( \alpha + 4\beta \right) 
  \right] \frac{r_h}{2G} \left( 1 - \frac{\Lambda}{3} r_h^2 \right), \label{E:tree}
\end{align}
which are exactly same with those obtained in ref.~\cite{Lu:2011zk}.
Note that  the factor $\left[ 1-2\alpha\Lambda/3 %1 + 2\Lambda \left( \alpha + 4\beta \right) 
\right]$ is vanishing at the critical point $\alpha %= -3\beta 
= 3/2\Lambda$. 
Thus, we can confirm that the energy and the entropy of the SAdS black hole at the critical point
are vanishing also in the Euclidean action formulation.

As was mentioned in the previous section, the entropy from the brick wall method  satisfies the area law and gives
the nontrivial thermodynamic quantities such as energy and heat capacity. 
At first glance, there seems to exist incompatibility between the Euclidean action formulation and the brick wall
method. In what follows, it will be shown that the semiclassical treatment of the Euclidean action formulation 
can be related to the brick wall method if we introduce the cutoff.

\section{Semiclassical  Euclidean action formulation}
\label{sec:pathintegral}

Now, we take the classical background as the SAdS black hole metric along with $\phi=0$, and
then consider the fluctuated quantum field semiclassically.
The partition function up to one loop order for the scalar field is expressed as
\begin{align}
  Z [g] &= Z^{(0)} [g] Z^{(1)} [g] \notag \\
  &= \exp \left( -\beta F^{(0)} \right) \exp \left( -\beta F^{(1)} \right) \notag \\
  &= e^{i I_\text{CG} [g]} \int \mathcal{D}\phi \, e^{i I_\phi [g,\phi]},
\end{align}
where the total free energy consists of $F = F^{(0)} + F^{(1)}$.
Note that the tree level free energy $F^{(0)}$ is trivial at the critical point as seen
in the previous section, so that the nontrivial contribution to the free
energy should come from the one loop effective action.

The one loop partition function $Z^{(1)}$ can be written as
\begin{align}
  Z^{(1)} [g] &= \int \mathcal{D}\phi \, e^{i I_\phi} \notag \\
  &= \det {}^{-1/2} (-\Box + m^2), 
\end{align}
and the effective action $W_\phi$ becomes
\begin{align}
 W_\phi &= \frac{i}{2} \ln \det (-\Box + m^2)
  \notag \\
  &= \frac{i}{2}\,  \mathrm{Tr} \ln  (-\Box + m^2) \notag \\
  &= \frac{i}{2} \int \frac{d^4 x d^4 k}{(2\pi)^4}\, \ln (k_\mu
  k^\mu + m^2), \label{I:x,k}
\end{align}
where $k_\mu$ corresponds to the conjugate momentum to $x^\mu$. 
Note that a (covariant) Fourier transform in curved spacetimes has not been 
established~\cite{Iagolnitzer:1991ic,Brunetti:1995rf,Brunetti:1999jn,Franco:2007ne}.
However, the manifold can be split into a number of small pieces, in which we can consider a
Riemann normal coordinates, i.e. $\int_\mathcal{M}d^4x \sqrt{-g} \simeq
\sum_{U\subset\mathcal{M}} \int_{U} d^4\tilde{x}$, where $\tilde{x}$ represents
 the Riemann normal coordinates~\cite{Birrell:1982ix}.
Then, one can perform the calculation in the momentum space by using the Fourier transform defined in 
this coordinates, $-\stackrel{\sim}{\Box}+m^2 \to \tilde{k}_\mu \tilde{k}^\mu + m^2$, where $\tilde{k}$ 
is the momentum measured in the local coordinates. 
Consequently, it is possible to recover the global coordinates back for the covariant result~\eqref{I:x,k}.

In the Euclidean geometry, the time component of the four vector $k_\mu$
becomes $2\pi n /(-i \beta)$, and the integrals $\int dt$ and $\int
dk_0/(2\pi)$ can be replaced by $-i \int d\tau$ and $(-i\beta)^{-1} \sum_n$, respectively~\cite{dj}. Then, the Euclidean
one loop effective action at the finite temperature is written as
\begin{align}
%  I_\phi &= - i 
  W_\phi &= \frac{1}{2} \sum_n \int \frac{d^3x d^3 k}{(2\pi)^3}
  \, \ln \left( \frac{4\pi^2 n^2}{f \beta^2} + E_m^2 \right), %\label{I:nsum}
  \notag \\
  &= \beta \int \frac{d^3x d^3 k}{(2\pi)^3} \left[
    \frac{\sqrt{f} E_m}{2} + \frac{1}{\beta} \ln \left( 1 -
      e^{-\beta \sqrt{f} E_m }\right) \right], \label{I:final}
\end{align}
where the relation $\sum_n \ln \left( \frac{4\pi^2 n^2}{\beta} + E_m^2 \right) = 2 \beta
  \left[ \frac{E_m}{2} + \frac{1}{\beta} \ln \left( 1- e^{-\beta E_m}\right) \right]$ was used
 and $E_m$ is defined by
\begin{equation}
  E_m^2 \equiv g^{ij} k_i k_j + m^2 = f k_r^2 + \frac{k_\theta^2}{r^2} +
  \frac{k_\phi^2}{r^2 \sin^2\theta} + m^2. \label{E:m:def}
\end{equation}
Actually,  the first term in the Euclidean one loop effective action~\eqref{I:final} 
is related to the vacuum energy of the spacetime which is independent of the black hole temperature
and we define the temperature dependent free energy as $\beta F^{(1)} = W_\phi - W_\phi^\text{vacuum}$.

Now, introducing a new parameter $\omega$, the free
energy can be rewritten as
\begin{align}
  F^{(1)} &= \frac{1}{\beta} \int d\omega \int \frac{d^3x d^3 k}{(2\pi)^3}
  \, \delta (\omega - \sqrt{f} E_m) \ln \left( 1 - e^{-\beta \omega}
  \right). \label{F:delta:omega}
\end{align}
Since the delta function can be written as the derivative of a step
function, $\delta(x) = \frac{d}{dx} \epsilon(x)$,
defining the step function as $\epsilon(x) = 1$ for
$x>0$ and $0$ for $x<0$,
then the free energy~\eqref{F:delta:omega} can be calculated as
\begin{align}
  F^{(1)} &= \frac{1}{\beta} \int \frac{d^3x d^3 k}{(2\pi)^3} \epsilon
  (\omega - \sqrt{f} E_m) \left. \ln \left( 1 - e^{-\beta \omega}
    \right) \right|_{-\infty}^\infty \notag \\
  &\quad - \int d^3 x \int_{m\sqrt{f}}^\infty d\omega\,
  \frac{1}{e^{\beta \omega} -1} \int_{V_p} \frac{d^3
    k}{(2\pi)^3} \label{F:part}
\end{align}
by performing the integration by parts.
$V_p$ is the
volume of the phase space satisfying $\sqrt{f} E_m \le \omega$,
which can be explicitly written as
\begin{align}
  f k_r^2 + \frac{k_\theta^2}{r^2} + \frac{k_\phi^2}{r^2 \sin^2\theta}
  \le \frac{\omega^2}{f} - m^2, \label{V:p}
\end{align}
for given values of $\omega$ and $m$. Subsequently, the
first term in eq.~\eqref{F:part} is vanishing, the free energy can be given by
\begin{align}
  F^{(1)} &= - \int d^3 x \int_{m\sqrt{f}}^\infty d\omega\,
  \frac{1}{e^{\beta \omega} -1} \int_{V_p} \frac{d^3
    k}{(2\pi)^3} \notag \\
  &= - \int_{0}^\infty d\omega\,
  \frac{1}{e^{\beta \omega} -1} \int_{V_p} \frac{d^3xd^3
    k}{(2\pi)^3}, \label{F:total}
\end{align}
where the relation $\int d^3x \int_{m\sqrt{f}}^\infty d\omega = \int_0^\infty d\omega \int_{\omega>m\sqrt{f}} d^3x$ was used.

Finally, the integration with respect to $x$ and $k$ can be replaced by the number of quantum states with energy less than $\omega$,
\begin{align}
  n(\omega) \equiv \int_{V_p} \frac{d^3x d^3 k}{(2\pi)^3} =
  \frac{1}{(2\pi)^3} \int_{V_p} dr d\theta d\phi dk_r dk_\theta
  dk_\phi, \label{n:omega}
\end{align}
then the free energy~\eqref{F:total} for a scalar field can be written as the remarkably familiar form of
\begin{align}
  F^{(1)} = - \int d\omega\, \frac{n(\omega)}{e^{\beta \omega} -1}. \label{F:final}
\end{align}
It is interesting to note that  the free energy~\eqref{F:final} obtained from the finite temperature one loop
effective action for a scalar field is the same with that given by the
brick wall method suggested by 't~Hooft~\cite{thooft} as it should be.

\section{Entropy of a Schwarzschild-anti de Sitter black hole}
\label{sec:thermo}

In this section, we will calculate the thermodynamic quantities of the SAdS black hole
using the free energy~\eqref{F:final}.  First, the number of quantum
states~\eqref{n:omega} with energy less than $\omega$ for a spherically symmetric
black hole is calculated as
\begin{align}
  n(\omega) &= \frac{2}{3\pi} \int dr \frac{ r^2 }{\sqrt{f}} \left(
      \frac{\omega^2}{f} - m^2 \right)^{3/2}. \label{n:omega:f}
\end{align}
Then, the free energy~\eqref{F:final} is obtained as
\begin{align}
  F^{(1)} &= - \frac{2}{3\pi} \int dr\, \frac{r^2}{f^2}
  \int_{m\sqrt{f}}^\infty d\omega\, \frac{\left(\omega^2 - m^2f
    \right)^{3/2}}{e^{\beta \omega}-1} \notag \\
  &= - \frac{2}{3\pi \beta^4} \int dr\, \frac{r^2}{f^2}
  \int_{z_0}^\infty \frac{dz}{e^z - 1}\left(z^2 - z_0^2 \right)^{3/2}, \label{F:drdomega}
\end{align}
where $z \equiv \beta \omega$ and $z_0 \equiv \beta m
\sqrt{f}$.
Subsequently, it is explicitly written as
\begin{equation}
  \label{F:epsilon}
  F^{(1)} = - \frac{2\pi^3 r_h^4}{45 \beta^4 \epsilon} (1-\Lambda
  r_h^2)^{-2}
\end{equation}
in the leading order where $r_h$, $\epsilon$ are the event horizon of the black hole and
the UV cutoff parameter. The UV cutoff $\epsilon$ is assumed to be very small
compared to the event horizon with a condition $m^2\ll r_+/[\epsilon \beta^2 (1-\Lambda r_+^2)]$.

Now, following 't~Hooft~\cite{thooft},  let us define the proper length for the UV cutoff parameter as
\begin{equation}
  \label{cutoff:length}
  \bar{\epsilon} \equiv \int_{r_h}^{r_h+\epsilon} dr \sqrt{g_{rr}} \approx
  \frac{2\sqrt{r_h \epsilon}}{\sqrt{1-\Lambda r_h^2}},
\end{equation}
which is independent of the parameters of the black hole.
Then, the free energy~\eqref{F:epsilon} is rewritten as
\begin{equation}
  \label{F:cutoff:length}
  F^{(1)} = - \frac{8\pi^3 r_h^5}{45 \beta^4 \bar\epsilon^2} (1-\Lambda
  r_h^2)^{-3}.
\end{equation}
The thermodynamic first law and the definition of the free energy gives
\begin{equation}
  \label{S:cutoff}
  S^{(1)} = \beta^2 \frac{\partial F^{(1)}}{\partial \beta} \bigg|_{\beta=\beta_H} = \frac{32 \pi^3
    r_h^5}{45 \beta_H^3 \bar\epsilon^2} (1-\Lambda
  r_h^2)^{-3},
\end{equation}
and the energy is obtained as
\begin{equation}
  \label{E:cutoff}
  E^{(1)} = F^{(1)} + \beta^{-1} S^{(1)} \big|_{\beta=\beta_H} = \frac{8\pi^3 r_h^5}{15 \beta_H^4 \bar\epsilon^2} (1-\Lambda
  r_h^2)^{-3}.
\end{equation}

Recovering dimensions and 
plugging the Hawking temperature~\eqref{T:H} into the
entropy~\eqref{S:cutoff}, one can get
\begin{align}
  S^{(1)} &= \frac{\ell_p^2}{90\pi \bar\epsilon^2} \frac{c^3\mathcal{A}}{4 G\hbar}, \label{S:final}
\end{align}
where $\mathcal{A} = 4\pi r_h^2$ and $\ell_p = \sqrt{G\hbar/c^3}$ are
the area of horizon and the Plank length, respectively.
The entropy~\eqref{S:final} agrees with the Bekenstein-Hawking entropy
$S^{(1)} = c^3 \mathcal{A} / (4G\hbar)$ when the cutoff is chosen
as $\bar\epsilon = \ell_p/\sqrt{90\pi}$, which is exactly same as in the
case of the Schwarzschild black hole~\cite{thooft}. Then, the free
energy~\eqref{F:cutoff:length} and the  energy~\eqref{E:cutoff} are rewritten as
\begin{align}
  F^{(1)} &= - \frac{c^4 r_h}{16 G}(1-\Lambda r_h^2) = - \frac{c^3
    \mathcal{A}}{16 G\hbar \beta_H}, \label{F:SAdS:final}
  \\
  E^{(1)} &= \frac{3c^4 r_h}{16 G}(1-\Lambda r_h^2) =
  \frac{3c^3\mathcal{A}}{16 G\hbar \beta_H}. \label{E:SAdS:final}
\end{align}
Next, the heat capacity is calculated as
\begin{align}
  C_V^{(1)} &\equiv T_H \frac{\partial S^{(1)}}{\partial T_H} = T_H \left(
    \frac{\partial S^{(1)}}{\partial r_h} \right) \left( \frac{\partial
      T_H}{\partial r_h} \right)^{-1} \notag \\
  &= - \frac{c^3 \mathcal{A}}{2 G \hbar} \left( \frac{1 - \Lambda r_h^2}{1 +
    \Lambda r_h^2} \right), \label{C:SAdS:final}
\end{align}
which is positive for $r_h > 1/ \sqrt{|\Lambda|}$ and negative
otherwise. It means that the SAdS black hole is stable for large black holes and unstable for small black holes, as is well-known. For $\Lambda=0$, the heat
capacity~\eqref{C:SAdS:final} is always negative, which coincides with the
thermodynamic stability of the Schwarzschild black hole.

%%%%%%%%%%%%%%%%%%%%%%%%%%%%%%%%%%%%%%%%%%%%%%%%%%%%%%%%%%%

% \begin{figure}[pt]
%   \includegraphics[width=0.5\textwidth]{metric}
%   \caption{The red-shift function $f(r)$ is shown with respect to
%     $r/\sqrt{\theta}$. There is no horizon for
%     $M=\sqrt{\theta}<M_0$(top), while one degenerate and two horizons
%     exist for $M=M_0\approx1.9\sqrt{\theta}$(middle) and
%     $M=3\sqrt{\theta}>M_0$(bottom), respectively.} 
%   \label{fig:metric}
% \end{figure}

\section{Discussion}
\label{sec:dis}
In summary, the four-dimensional critical gravity seems to be trivial in that the entropy of
the SAdS black hole is vanishing, which may be a somewhat impatient conclusion because the
black hole temperature is not zero which means that the Hawking radiation exists. 
In the Euclidean action formulation at finite temperature,
the total free energy consists of the tree free energy and the one loop corrected
free energy $ F = F^{(0)} + F^{(1)}$, which yields the total entropy $S = S^{(0)} + S^{(1)}$.
In spite of the vanishing tree entropy $ S^{(0)}=0$, the total entropy gives
the area law by taking into account the quantum
fluctuation of the scalar field. 

So far, we have assumed that
 both the scalar decoupling condition and the critical condition are valid since the metric field has been fixed. 
Now, one may wonder whether these conditions are still met or not when one considers 
 the one loop back reaction of the geometry because it may affect the thermodynamic quantities. 
For this purpose, the coefficients in front of higher curvature terms can be released as
$I[g] = \frac{1}{16\pi G_B} \int d^4x \sqrt{-g} \left[ R - 2\Lambda_B + \alpha_B R_{\mu\nu} R^{\mu\nu} + \beta_B R^2 \right]$, where the parameters
are bare couplings defined by, e.g., $\alpha_B = \alpha + \hbar \delta \alpha$.
Of course, it recovers the critical gravity for the particular choice of
$\alpha + 3\beta = 0$ and $\alpha - 3/2\Lambda = 0$ at the tree level.
Now, the one loop effective action of the massive scalar field whose mass is $m$ can be 
also written in the form of the divergent higher curvature terms so that the renormalization yields     
$\alpha_R/G_R = \alpha_B/G_B + \hbar A/120\pi$ and $\beta_R/G_R = \beta_B/G_B + \hbar A/240\pi$, %$G_R = G_B - \hbar G_B^2 B/12\pi$
where $\alpha_R$, $\beta_R$ and $G_R$ are
the renormalized couplings. Note that $A \approx \ln (\mu^2/m^2) + 2\ln(2/3) + O(m^2/\mu^2)$
is a divergent constant for $\mu\to\infty$~\cite{Birrell:1982ix,Demers:1995dq}.
It means that $\alpha_R + 3\beta_R =(\hbar G_N/24\pi) \ln(2/3)$ and
$\alpha_R - 3/2\Lambda_R = (\hbar G_N/480\pi \Lambda^2) [ 4( 2\Lambda^2 + 60 \Lambda m^2 - 45 m^4) \ln(2/3) + 15 m^2 ( 8\Lambda - 9m^2 ) ]$
by appropriate counter terms. Fortunately, by rescaling $\mu\to3\mu/2$, one can 
still require $\alpha_R + 3\beta_R = 0$ in order to avoid the existence of the scalar graviton; 
 however, the critical condition is still violated as
$\alpha_R - 3/2\Lambda_R = (\hbar G_N/32\pi \Lambda^2) m^2 ( 8\Lambda - 9m^2 )$. 
Therefore, one cannot maintain the scalar graviton decoupling condition and the critical condition simultaneously 
on account of the back  reaction of the geometry. But, further study is needed for 
the concrete argument and relevant thermodynamic quantities in connection
with the back reaction of the geometry. We hope this issue will be addressed elsewhere.

%\section*{Acknowledgments}
\acknowledgments 
W.~Kim would like to thank H.  Lu for exciting discussion for the critical gravity.
 E.~J.~Son were supported by the National Research Foundation of Korea(NRF) grant
funded by the Korea government(MEST) through the Center for Quantum
Spacetime(CQUeST) of Sogang University with grant number 2005-0049409.
M.~Eune was supported by National Research Foundation
of Korea Grant funded by the Korean Government (Ministry of Education,
Science and Technology) (NRF-2010-359-C00007).
W.~Kim was  supported by the Basic Science Research Program
through the National Research Foundation of Korea(NRF) funded by the
Ministry of Education, Science and Technology(2010-0008359).

%%%%%%%%%%%%%%%%%%%%%%%%%%%%%%%%%%%%%%%%%%%%%%%%%%%%%%%%%%%%%
%%%%%%%%%%%%%%%             References       %%%%%%%%%%%%%%%%
%%%%%%%%%%%%%%%%%%%%%%%%%%%%%%%%%%%%%%%%%%%%%%%%%%%%%%%%%%%%%


\begin{thebibliography}{99}

%\cite{Stelle:1976gc}
\bibitem{Stelle:1976gc}
  K.~S.~Stelle,
  \textit{Renormalization of Higher Derivative Quantum Gravity,}
  Phys.\ Rev.\ {\bf D16} (1977) 953-969.

%\cite{Stelle:1977ry}
\bibitem{Stelle:1977ry}
  K.~S.~Stelle,
  \textit{Classical Gravity with Higher Derivatives,}
  Gen.\ Rel.\ Grav.\ {\bf 9} (1978) 353-371.
  
%\cite{Deser:1981wh}
\bibitem{Deser:1981wh}
  S.~Deser, R.~Jackiw, S.~Templeton,
  \textit{Topologically Massive Gauge Theories,}
  Ann.\ Phys.\ (N.Y.) {\bf 140} (1982) 372-411.

%\cite{Deser:1982vy}
\bibitem{Deser:1982vy}
  S.~Deser, R.~Jackiw, S.~Templeton,
  \textit{Three-Dimensional Massive Gauge Theories,}
  Phys.\ Rev.\ Lett.\ {\bf 48} (1982) 975-978.
  
%\cite{Li:2008dq}
\bibitem{Li:2008dq}
  W.~Li, W.~Song and A.~Strominger,
  \textit{Chiral Gravity in Three Dimensions,}
  JHEP {\bf 0804} (2008) 082.
  [arXiv:0801.4566 [hep-th]].

%\cite{Lu:2011zk}
\bibitem{Lu:2011zk}
  H.~Lu, C.~N.~Pope,
  \textit{Critical Gravity in Four Dimensions,}
  Phys.\ Rev.\ Lett.\ {\bf 106} (2011) 181302.
  [arXiv:1101.1971 [hep-th]].


%\cite{Gibbons:1976ue}
\bibitem{Gibbons:1976ue}
  G.~W.~Gibbons, S.~W.~Hawking,
  \textit{Action Integrals and Partition Functions in Quantum Gravity,}
  Phys.\ Rev.\ {\bf D15} (1977) 2752-2756.

%\cite{Gibbons:1978ac}
\bibitem{Gibbons:1978ac}
  G.~W.~Gibbons, S.~W.~Hawking, M.~J.~Perry,
  \textit{Path Integrals and the Indefiniteness of the Gravitational Action,}
  Nucl.\ Phys.\ {\bf B138} (1978) 141.

%\cite{Hawking:1982dh}
\bibitem{Hawking:1982dh}
  S.~W.~Hawking, D.~N.~Page,
  \textit{Thermodynamics of Black Holes in anti-De Sitter Space,}
  Commun.\ Math.\ Phys.\ {\bf 87} (1983) 577.

%\cite{Bekenstein:1972tm}
\bibitem{Bekenstein:1972tm}
  J.~D.~Bekenstein,
  \textit{Black holes and the second law,}
  Lett.\ Nuovo Cim.\ {\bf 4} (1972) 737-740.

%\cite{Bekenstein:1973ur}
\bibitem{Bekenstein:1973ur}
  J.~D.~Bekenstein,
  \textit{Black holes and entropy,}
  Phys.\ Rev.\ {\bf D7} (1973) 2333-2346.

%\cite{Bekenstein:1974ax}
\bibitem{Bekenstein:1974ax}
  J.~D.~Bekenstein,
  \textit{Generalized second law of thermodynamics in black hole physics,}
  Phys.\ Rev.\ {\bf D9} (1974) 3292-3300.

%\cite{Hawking:1974sw}
\bibitem{Hawking:1974sw}
  S.~W.~Hawking,
  \textit{Particle Creation by Black Holes,}
  Commun.\ Math.\ Phys.\ {\bf 43} (1975) 199-220.

\bibitem{thooft}
  G.~'t~Hooft,
  \textit{On the Quantum Structure of a Black Hole,}
  Nucl.\ Phys.\ {\bf B256} (1985) 727.
  %%CITATION = NUPHA,B256,727;%%

%\cite{Hohm:2010jc}
\bibitem{Hohm:2010jc}
  O.~Hohm, E.~Tonni,
  \textit{A boundary stress tensor for higher-derivative gravity in AdS and Lifshitz backgrounds,}
  JHEP {\bf 1004} (2010) 093.
  [arXiv:1001.3598 [hep-th]].



%% microlocal analysis
%\cite{Iagolnitzer:1991ic}
\bibitem{Iagolnitzer:1991ic}
  D.~Iagolnitzer,
  \textit{Microlocal analysis and phase space decompositions,}
  Lett.\ Math.\ Phys.\ {\bf 21} (1991) 323-328.

%\cite{Brunetti:1995rf}
\bibitem{Brunetti:1995rf}
  R.~Brunetti, K.~Fredenhagen, M.~Kohler,
  \textit{The Microlocal spectrum condition and Wick polynomials of free fields on curved space-times,}
  Commun.\ Math.\ Phys.\ {\bf 180} (1996) 633-652.
  [gr-qc/9510056].

%\cite{Brunetti:1999jn}
\bibitem{Brunetti:1999jn}
  R.~Brunetti, K.~Fredenhagen,
  \textit{Microlocal analysis and interacting quantum field theories: Renormalization on physical backgrounds,}
  Commun.\ Math.\ Phys.\ {\bf 208} (2000) 623-661.
  [math-ph/9903028].

%\cite{Franco:2007ne}
\bibitem{Franco:2007ne}
  D.~H.~T.~Franco, J.~L.~Acebal,
  \textit{Microlocal analysis and renormalization in finite temperature field theory,}
  Int.\ J.\ Theor.\ Phys.\ {\bf 46} (2007) 383-398.
  [hep-th/0306169].


%\cite{Birrell:1982ix}
\bibitem{Birrell:1982ix}
  N.~D.~Birrell and P.~C.~W.~Davies,
  \textit{Quantum Fields In Curved Space,}
  Cambridge, Uk: Univ. Pr. (1982) 340p.

\bibitem{dj}
  L.~Dolan and R.~Jackiw,
  \textit{Symmetry Behavior at Finite Temperature,}
  Phys.\ Rev.\ {\bf D9} (1974) 3320.
  %%CITATION = PHRVA,D9,3320;%%

%\cite{Lu:2011ks}
\bibitem{Lu:2011ks}
  H.~Lu, Y.~Pang and C.~N.~Pope,
  \textit{Conformal Gravity and Extensions of Critical Gravity,}
  %
  [arXiv:1106.4657 [hep-th]].

%\cite{Demers:1995dq}
\bibitem{Demers:1995dq}
  J.~-G.~Demers, R.~Lafrance and R.~C.~Myers,
  \textit{Black hole entropy without brick walls,}
  Phys.\ Rev.\  D {\bf 52} (1995)  2245-2253.
  [gr-qc/9503003].


\end{thebibliography}
\end{document}